\documentclass[12pt]{article}
\usepackage[dvipdfmx]{graphicx}
\usepackage{amsmath,amssymb,epsfig,cite,latexsym}
\allowdisplaybreaks[3]

\begin{document}
\rightline{December, 2014}

\vspace{25mm}
\begin{center}
{\Large\bf Brane induced cosmological acceleration and\\
\vspace{0.5\baselineskip}
crossing of $w_{eff}=-1$}

\vspace{20mm}
\def\thefootnote{\fnsymbol{footnote}}
{\large Nobuyuki MOTOYUI\footnote[1]{motoyui@mx.ibaraki.ac.jp}}

\vspace{20mm}
{\large
Department of Physics, Faculty of Sciences,\\
\vspace{0.5\baselineskip}
Ibaraki University, Bunkyo 2-1-1, Mito, 310-8512, Japan}
\end{center}

\vspace{15mm}
\begin{abstract}
The cosmological observation indicates that the effective equation
of state parameter $w_{eff}$ varies with $z$:
It changes from $w_{eff}>-1$ to $w_{eff}<-1$ at $z\sim 0.2$.
We investigate under which condition it exhibits such behaviors 
based on the five-dimensional braneworld scenario.
It is possible in the model with or without an energy exchange between
the four dimensional universe and the fifth dimension.
However the curves of $w_{eff}$ are quite different between the two cases.
\end{abstract}

\vspace{5mm}
\noindent
Keywords: Brane universe; dark energy;
effective equation of state parameter.
\\
PACS numbers: 95.36.+x, 11.25.Mj.

\newpage
\section{Introduction}

The cosmological observations indicate that our universe have not only
experienced the large amount of accelerating expansion in its infancy
but it is still undergoing a small rate of accelerating
expansion~\cite{Riess,Perl}.
The present cosmological data also indicates that more than
70\% of the total energy density is attributed to a component of
dark energy density~\cite{WMAP9-a,WMAP9-b}.
The source of driving force to the present accelerating expansion is
thought to be dark energy.
Although the physical origin of dark energy is still unknown,
many cosmological candidates such as the cosmological constant or
several kinds exotic matter like phantom fields~\cite{Phantom-a,Phantom-b},
quintessence~\cite{Quint-a,Quint-b,Quint-c},
or modifications of gravitational theory have been proposed.
If we define the ratio of the pressure of the universe $P$ to
its energy density $\rho$ as $w \equiv P/\rho$,
then the cosmological constant is characterized by $w=-1$ and
the phantom field is characterized by $w<-1$. If the cosmological constant
is identified as dark energy, $w$ is a constant.
On the other hand, $w$ varies with time if we adopt the assumptions of
exotic matter. Although the simplest candidate of dark energy is
the cosmological constant $\Lambda$,
there is a well known 'fine tuning problem' between the energy density of
cosmological constant $\rho_{\Lambda}$ and the radiation density $\rho_{r}$:
The ratio of $\rho_{\Lambda}$ and $\rho_{r}$ depends on the energy scale and
$\rho_{\Lambda}/\rho_{r}\sim 10^{-123}$ at the Planck scale, on the other
hand $\rho_{\Lambda}/\rho_{r}\sim 10^{-54}$ at the electroweak scale
because $\rho_{\Lambda}\simeq 10^{-47} GeV^{4}$ is a constant.

However the effective equation of state parameter is not necessarily a constant.
Alternative ways for dark energy model in which either dark energy or its
effective equation of state parameter is a function of time.
The cosmological data indicates that the time varying dark energy model is
a better fitting than a cosmological constant:
the effective equation of state parameter of dark energy
$w_{eff}\sim -1.21$ at $z=0$ and it changes
from $w_{eff}>-1$ to $w_{eff}<-1$ at $z\sim 0.2$~\cite{Alam}.
It indicates that $w_{eff}$ behaves as phantom-like at lower redshifts
$z\lesssim 0.2$ and dust-like at higher redshifts.
Since a phantom component with $w_{eff}<-1$ violates energy conditions,
one may hope to make a dark energy model with the above nature in
the framework of effective features, not caused by a phantom field. Then
we may hope to do this behavior in the framework of braneworld scenario.

In the braneworld scenario,
our universe is realized as a three dimensional hypersurface (brane) which
is embedded in a higher dimensional spacetime (bulk).
Gravity can propagate to the entire spacetime while the ordinary matter is
confined on the brane.
It received much attentions due to the possibilities that
the compactification with the large extra dimensions~\cite{Arkani},
the solution to the large disparity between the electroweak scale and
the Planck scale~\cite{RS1}.
It is an important question that whether the standard four dimensional
gravity is reproduced in these scenario. It is showed that the massless
gravitons are trapped on the brane and the four dimensional gravity is
reproduced~\cite{RS2}.

It is an interesting question that how the present accelerating expansion
of our universe can be treated in the framework of the braneworld scenario.
The evolution of our universe is described as an effective Friedmann equation
on a brane. It is modified from the usual Friedmann equation because of
the presence of the fifth dimension~\cite{Binetruy-a,Binetruy-b}.
The presence of the fifth dimension allows a five-dimensional bulk matter
which propagates in the fifth dimension. It may interact with the ordinary
matter on the brane and it could be able to lead the behavior that
resembles to dark energy.
In this picture, the bulk pressure and the off-diagonal terms in
the energy-momentum tensor affect the cosmological evolution of the brane.
It has been shown that $w_{eff}$ crosses $w_{eff}=-1$ at lower redshift
in the model which allows the energy exchange between the four dimensional
universe and the fifth dimension~\cite{Cai,Bogdanos-a,Bogdanos-b}.

This paper is organized as follows. In section 2, we present the general
framework of the five-dimensional theory. Then we present the effective
Friedmann equation and the conservation of energy-momentum
based on \cite{Binetruy-a}.
In section 3, we derive the Hubble equation on the 3-brane.
It is quite different from that of a standard four dimensional
Friedmann-Lema\^{i}tre-Robertson-Walker (FLRW) cosmology in
several points. Then we derive the effective equation of state parameter
$w_{eff}$ of dark energy.
In section 4, we examine the evolutions of $w_{eff}$ and investigate under
which condition it exhibits crossing of $w_{eff}=-1$ at $z\sim 0.2$
and $w_{eff}\sim -1.21$ at $z=0$ in two cases:
(i) there is energy exchange between the four dimensional
universe and the fifth dimension, and
(ii) there is no energy exchange between the four dimensional universe
and the fifth dimension.
We also examine the evolutions of deceleration parameter
and energy density components. Section 5 is devoted to summary.

\section{General Framework and the effective Friedmann equation}

We consider the five-dimensional action of the following form,
\begin{eqnarray}
S=
\int d^{5}x\sqrt{-g_{(5)}} \left[
\frac{1}{2\kappa_{(5)}^{2}}\left(R_{(5)}-2\Lambda_{(5)}\right)
+\mathcal{L}_{B}^{(m)}
\right]
+\int d^{5}x\,\sqrt{-g_{b}} \left(-T_{b}+\mathcal{L}_{b}^{(m)}
\right)\delta(y).
\label{eq:5D1}
\end{eqnarray}

\noindent
We denote the coordinate of fifth dimension by $y$ and it takes the values
$-\infty <y<\infty$. We consider a 3-brane is located at $y=0$ and
assume a $Z_{2}$ symmetry for $y$ around $y=0$.
In the above expression,
$R_{(5)}$ is the five-dimensional Ricci scalar, $\Lambda_{(5)}$
is the five-dimensional cosmological constant and $T_{b}$ is the tension
of the 3-brane. We can include matter content in the bulk
$\mathcal{L}_{B}^{(m)}$ or on the brane $\mathcal{L}_{b}^{(m)}$.
We denote the five-dimensional metric as $g_{(5)AB}$,
the four dimensional metric as $g_{\mu\nu}$ and
the four dimensional metric on the brane as $g_{b}$.
We define the signature of $g_{(5)AB}$ as $(-,+,+,+,+)$ and that of
$g_{\mu\nu}$ as $(-,+,+,+)$. The line element is described as
\begin{eqnarray}
ds^{2}=g_{(5)AB}dx^{A}dx^{B}=g_{\mu\nu}dx^{\mu}dx^{\nu}+b^{2}(t,y)dy^{2}.
\label{eq:Met1}
\end{eqnarray}

\noindent
The capital Latin indices indicate ($0, \cdots, 4$)
and the Greek indices indicate ($0, \cdots, 3$).
The constant $\kappa_{(5)}$ is related to the five-dimensional
Newton constant $G_{(5)}$ and the five-dimensional Planck mass $M_{(5)}$
by the relation $\kappa_{(5)}^{2}=8\pi G_{(5)}=M_{(5)}^{-3}$.
We assume that the five-dimensional metric is described as follows,
\begin{eqnarray}
ds^{2}=
-n^{2}(t,y)dt^{2}+a^{2}(t,y)\gamma_{ij}dx^{i}dx^{j}+b^{2}(t,y)dy^{2},
\label{eq:Met2}
\end{eqnarray}

\noindent
where $\gamma_{ij}$ is a maximally symmetric FLRW metric. Its spatial
curvature is parametrized by $K$ which takes the values $K=-1,0,1$.

The energy-momentum tensor $T_{AB}$ is given as
\begin{eqnarray}
T_{AB}&=&
T^{(B)}{}_{AB}+T^{(b)}{}_{AB}
-T_{b}\frac{\sqrt{-g_{b}}}{\sqrt{-g_{(5)}}}\,
g_{\mu\nu}\delta^{\mu}_{A}\delta^{\nu}_{B}\delta (y),
\label{eq:Em1}
\end{eqnarray}

\noindent
where $T^{(B)}{}_{AB}$ is the component which results from
$\mathcal{L}_{B}^{(m)}$ and $T^{(b)}{}_{AB}$ is the component which
results from $\mathcal{L}_{b}^{(m)}$.
We assume the bulk energy-momentum tensor $T^{(B)}{}^{A}{}_{B}$ as
\begin{eqnarray}
T^{(B)}{}^{A}{}_{B}&=&
\left(
  \begin{array}{ccc}
    -\rho_{B} &  0  &  Q  \\
      0 & P_{B}\delta^{i}{}_{j} &  0  \\
    -\frac{n^{2}}{b^{2}}Q & 0 & P_{T} \\
  \end{array}
\right).
\label{eq:Em2}
\end{eqnarray}

\noindent
The brane energy-momentum tensor $T^{(b)}{}^{A}{}_{B}$ is generally
expressed as
\begin{eqnarray}
T^{(b)}{}^{A}{}_{B}&=&
\frac{\delta(y)}{b}
\left(
  \begin{array}{ccc}
    -\rho_{b} &  0  &  0  \\
      0  & P_{b}\delta^{i}{}_{j} &  0  \\
      0  &  0  &  0  \\
  \end{array}
\right).
\label{eq:Em3}
\end{eqnarray}

\noindent
In the above expression, $Q$ is responsible for the energy exchange
between the four dimensional spacetime and the extra dimension.
We generally allow the anisotropic choice of the bulk pressure:
$P_{B}\neq P_{T}$.
The dynamics of the five-dimensional universe is governed by
the five-dimensional Einstein equations. They take the usual form,
\begin{eqnarray}
G_{(5)AB}\equiv R_{(5)AB}-\frac{1}{2}g_{(5)AB}R_{(5)}
=-\Lambda_{(5)}\, g_{AB} +\kappa_{(5)}^{2}T_{AB}.
\label{eq:Ee1}
\end{eqnarray}

\noindent
Substituting (\ref{eq:Met2}), (\ref{eq:Em2}) and (\ref{eq:Em3}) into
(\ref{eq:Ee1}), we obtain the following equations,
\begin{eqnarray}
&&3\left\{
\frac{\dot{a}}{a}\left(\frac{\dot{a}}{a}+\frac{\dot{b}}{b}\right)
-\frac{n^{2}}{b^{2}}\left(
\frac{a''}{a}+\frac{a'}{a}\left(\frac{a'}{a}-\frac{b'}{b}\right)\right)
+K\frac{n^{2}}{a^{2}} \right\}
\nonumber\\
&&~~~~=~
n^{2}\left(\Lambda_{(5)}+\kappa_{(5)}^{2}\rho_{B}
+\frac{\kappa_{(5)}^{2}}{b}(\rho_{b}+T_{b})\delta(y)\right),
\label{eq:Ee3}\\
&&\frac{a^{2}}{b^{2}}\left\{
\frac{a'}{a}\left(\frac{a'}{a}+2\frac{n'}{n}\right)
-\frac{b'}{b}\left(\frac{n'}{n}+2\frac{a'}{b}\right)
+2\frac{a''}{a}+\frac{n''}{n}
\right\}
\nonumber\\
&&+\frac{a^{2}}{n^{2}}\left\{
\frac{\dot{a}}{a}\left(-\frac{\dot{a}}{a}+2\frac{\dot{n}}{n}\right)
-2\frac{\ddot{a}}{a}
+2\frac{\dot{b}}{b}\left(-\frac{\dot{a}}{a}+\frac{\dot{n}}{n}\right)
-\frac{\ddot{b}}{b}
\right\}
+K
\nonumber\\
&&~~~~=~
a^{2}\left(-\Lambda_{(5)}+\kappa_{(5)}^{2}P_{B}
+\frac{\kappa_{(5)}^{2}}{b}(P_{b}-T_{b})\delta(y)\right),
~~~~~
\label{eq:Ee4}\\
&&3\left(\frac{n'}{n}\frac{\dot{a}}{a}
+\frac{a'}{a}\frac{\dot{b}}{b}-\frac{\dot{a}'}{a}\right)
~=~
-n^{2}\kappa_{(5)}^{2}Q,
\label{eq:Ee5}\\
&&3\left\{
\frac{a'}{a} \left(\frac{a'}{a}+\frac{n'}{n}\right)
-\frac{b^{2}}{n^{2}} \left(
\frac{\dot{a}}{a} \left(\frac{\dot{a}}{a}-\frac{\dot{n}}{n}\right)
+\frac{\ddot{a}}{a} \right)
-K\frac{b^{2}}{a^{2}}
\right\}
~=~
b^{2}\left(-\Lambda_{(5)}+\kappa_{(5)}^{2}P_{T}\right),~~~~
\label{eq:Ee6}
\end{eqnarray}

\noindent
where dots stand for differentiations with respect to $t$ and
primes stand for differentiations with respect to $y$.

We make two assumptions to simplify these equations:

\vspace*{-0.25\baselineskip}
{\newcounter{Ai}
\newcommand{\Ai}{\refstepcounter{Ai}(\roman{Ai})}
\begin{list}{\Ai}{\itemsep=4pt \parsep=0pt \leftmargin=2em}
\item
the scale factor of extra dimension is normalized to $b(t,y)=1$,

\item
the scale factor of temporal dimension at $y=0$ is normalized to
$n(t,0)=1$.
\end{list}}

\noindent
After these simplifications, Einstein equations on the brane are expressed
as follows,
\begin{eqnarray}
&&3\left\{
\left(\frac{\dot{a}_{b}}{a_{b}}\right)^{2}
-\frac{a_{b}''}{a_{b}} -\left(\frac{a_{b}'}{a_{b}}\right)^{2}
+K\frac{1}{a_{b}^{2}} \right\}
~=~
\Lambda_{(5)}+\kappa_{(5)}^{2}\rho_{B}
+\kappa_{(5)}^{2}(\rho_{b}+T_{b})\delta(y),
\label{eq:Eeb1}\\
&&\left\{
\frac{a_{b}'}{a_{b}}\left(\frac{a_{b}'}{a_{b}}+2{n_{b}'}\right)
+2\frac{a_{b}''}{a_{b}}+{n_{b}''}
\right\}
-\left(\frac{\dot{a}_{b}}{a_{b}}\right)^{2}
-2\frac{\ddot{a}_{b}}{a_{b}}
-K\frac{1}{a_{b}^{2}}
\nonumber\\*
&&~~~~=~
-\Lambda_{(5)}+\kappa_{(5)}^{2}P_{B}+\kappa_{(5)}^{2}(P_{b}-T_{b})\delta(y),
\label{eq:Eeb2}\\
&&3\left({n_{b}'}\frac{\dot{a}_{b}}{a_{b}}
-\frac{\dot{a}_{b}'}{a_{b}}\right)
~=~
-\kappa_{(5)}^{2}Q,
\label{eq:Eeb3}\\
&&3\left\{
\frac{a_{b}'}{a_{b}} \left(\frac{a_{b}'}{a_{b}}+{n_{b}'}\right)
-\left(\frac{\dot{a}_{b}}{a_{b}}\right)^{2}
-\frac{\ddot{a}_{b}}{a_{b}}
-K\frac{1}{a_{b}^{2}}
\right\}
~=~
-\Lambda_{(5)}+\kappa_{(5)}^{2}P_{T},
\label{eq:Eeb4}
\end{eqnarray}

\noindent
where the subscripts $b$ denotes that these functions are evaluated at $y=0$.

We have to take into account of the junction conditions~\cite{Binetruy-a}
when we solve the Einstein equations on the brane.
Because the first derivatives of the metric with respect to $y$ can be
discontinuous at $y=0$ while the metric is required to be continuous across
the brane, the delta functions appear in the second derivatives of
the metric. We define the jump factor $\ddag f\ddag$
and the mean value $\sharp f\sharp$ of a function $f$ by the following
equations,
\begin{eqnarray}
\ddag f\ddag\equiv f(t,0+)-f(t,0-),
\label{eq:Jc1}\\
\sharp f\sharp\equiv \frac{f(t,0+)+f(t,0-)}{2}.
\label{eq:Jc2}
\end{eqnarray}

\noindent
We obtain the junction conditions as follows,
\begin{eqnarray}
\frac{\ddag a'\ddag}{a_{b}}
&=&-\frac{\kappa_{(5)}^{2}}{3}(\rho_{b}+T_{b})b_{b},
\label{eq:Jc9}\\
\frac{\ddag n'\ddag}{n_{b}}
&=&\frac{\kappa_{(5)}^{2}}{3}(2\rho_{b}+3P_{b}-T_{b})b_{b}.
\label{eq:Jc10}
\end{eqnarray}

\noindent
Taking into account the $Z_{2}$ symmetry around $y=0$, we obtain
\begin{eqnarray}
\frac{a'(t,+0)}{a_{b}}
&=&-\frac{\kappa_{(5)}^{2}}{6}(\rho_{b}+T_{b}),
\label{eq:Ec8}\\
{n'(t,+0)}
&=&\frac{\kappa_{(5)}^{2}}{6}(2\rho_{b}+3P_{b}-T_{b}).
\label{eq:Ec9}
\end{eqnarray}

\noindent
Taking the jump of (\ref{eq:Eeb4}), we also obtain a relation between
mean values $\sharp a'\sharp$ and $\sharp n'\sharp$,

\vspace*{-0.75\baselineskip}
\begin{eqnarray}
\frac{\sharp a'\sharp}{a_{b}}(P_{b}-T_{b})
&=&
\frac{1}{3}(\rho_{b}+T_{b}) \frac{\sharp n'\sharp}{n_{b}}.
\label{eq:Jc11}
\end{eqnarray}

\noindent
Substituting (\ref{eq:Ec8}) and (\ref{eq:Ec9}) in (\ref{eq:Eeb4}),
we obtain the effective Friedmann equation in the limit $y\to +0$,
\begin{eqnarray}
\frac{\ddot{a}_{b}}{a_{b}} +\left(\frac{\dot{a}_{b}}{a_{b}}\right)^{2}
+\frac{K}{a_{b}^{2}}
&=&-\frac{\kappa_{(5)}^{4}}{36}(1+3w_{b})\rho_{b}^{2}
+\frac{\kappa_{(5)}^{4}}{36}(1-3w_{b})T_{b}\rho_{b}
\nonumber\\
&&+\frac{1}{3}\left(\Lambda_{(5)}+\frac{\kappa_{(5)}^{4}}{6}T_{b}^{2}\right)
-\frac{\kappa_{(5)}^{2}}{3}P_{T}.
\label{eq:Eeb5}
\end{eqnarray}

\noindent
In the above expression, we used the equation of state for the brane:
$P_{b}=w_{b}\rho_{b}$.
Note that the 55-component of the bulk energy-momentum tensor $P_{T}$
and the quadratic term of the brane energy density $\rho_{b}$ appear
on the right hand side of this equation. They will affect the cosmological
evolution of the 3-brane.

The equations of Energy-momentum conservation are,
\begin{eqnarray}
\nabla_{A}T^{A}_{B}=
\partial_{A}T^{A}_{B}+\Gamma^{A}_{DA}T^{D}_{B}-\Gamma^{D}_{BA}T^{A}_{D}=0.
\label{eq:Ec1}
\end{eqnarray}

\noindent
The 0-component and the 5-component of the above equations are
\begin{eqnarray}
\dot{\rho}_{B}
+\frac{n^{2}}{b^{2}}Q'
+3(P_{B}+\rho_{B})\frac{\dot{a}}{a}
+3Q\frac{n^{2}}{b^{2}}\left(\frac{n'}{n}+\frac{a'}{a}\right)
-Q\frac{n^{2}b'}{b^{3}}&&
\nonumber\\
+(P_{T}+\rho_{B})\frac{\dot{b}}{b}
+\delta(y)\left\{
\dot{\rho}_{b} +3(P_{b}+\rho_{b})\frac{\dot{a}}{a}
+\rho_{b}\frac{\dot{b}}{b}
\right\}
&=&0,
\label{eq:Ec2}\\
\dot{Q}
+{P_{T}}'
+Q\left(\frac{\dot{n}}{n}+3\frac{\dot{a}}{a}+\frac{\dot{b}}{b}\right)
+(P_{T}+\rho_{B})\frac{n'}{n}
+3(P_{T}-P_{B})\frac{a'}{a}
&=&0.
\label{eq:Ec3}
\end{eqnarray}

\noindent
Integrating (\ref{eq:Ec2}) around $y=0$ and using the $Z_{2}$ symmetry,
we obtain the conservation of energy-momentum on the brane,
\begin{eqnarray}
\dot{\rho}_{b} +3(1+w_{b})\rho_{b}\frac{\dot{a}_{b}}{a_{b}}
+2Q(t)
&=&0,
\label{eq:Ebr1}
\end{eqnarray}

\noindent
where we used two assumptions  $b(t,y)=1$ and $n(t,0)=1$.
Using (\ref{eq:Ec8}) and (\ref{eq:Ec9}), we obtain
the conservation of energy-momentum in the limit $y\to +0$,
\begin{eqnarray}
\dot{\rho}_{B}
+Q'
+3(P_{B}+\rho_{B})\frac{\dot{a}_{b}}{a_{b}}
+\frac{\kappa_{(5)}^{2}}{2}Q \left\{ (1+3w_{b})\rho_{b}-2T_{b} \right\}
=0,
\label{eq:Ebu1}\\
\dot{Q}
+{P_{T}}'
+3Q\frac{\dot{a}_{b}}{a_{b}}
+\frac{\kappa_{(5)}^{2}}{6}(P_{T}+\rho_{B})
\left\{ (2+3w_{b})\rho_{b}-T_{b} \right\}
-\frac{\kappa_{(5)}^{2}}{2}(P_{T}-P_{B})(\rho_{b}+T_{b})
=0,
\label{eq:Ebu2}
\end{eqnarray}

\noindent
where we used the two assumptions for the scale factors of
extra dimension and the temporal dimension.
We obtained an effective Friedmann equation on the 3-brane,
the evolution equations for $\rho_{b}$, $\rho_{B}$ and $Q$.

The time evolution of the bulk energy density $\rho_{B}$ and the flow of
energy from/to the extra dimension $Q$ are governed by (\ref{eq:Ebu1})
and (\ref{eq:Ebu2}). The bulk pressure $P_{T}$ contributes to
the effective Friedmann equation on a brane through (\ref{eq:Eeb5}).
However we cannot fully determine $Q$ and $P_{T}$ since there are unknown
functions $Q'$ and $P_{T}'$ in (\ref{eq:Ebu1}) and (\ref{eq:Ebu2}).
We put two ans\"{a}tze for $Q$ and $P_{T}$ as follows,
\begin{eqnarray}
Q=F\left(\frac{\dot{a}_{b}(t)}{a_{b}(t)}\right) a_{b}(t)^{\mu},~~
P_{T}=Da_{b}(t)^{\nu},
\label{eq:Chi7}
\end{eqnarray}

\noindent
where $D,\, F,\, \mu$ and $\nu$ are some constants.
A justification of these ans\"{a}tze is found in~\cite{Bogdanos-a}.

\section{Hubble equation and effective equation of state}

To transform the second order equation (\ref{eq:Eeb5}) to the first order
equation, we introduce a new variable $\chi(t)$ which is called
{\it dark energy variable}~\cite{Bogdanos-a},
\begin{eqnarray}
\chi
&\equiv&
\left(\frac{\dot{a}_{b}}{a_{b}}\right)^{2} +\frac{K}{a_{b}^{2}}
-2\gamma\rho_{b} -\beta\rho_{b}^{2} -\frac{\lambda}{2}
+\frac{\kappa_{(5)}^{2}}{6}P_{T},
\label{eq:Chi1}
\end{eqnarray}

\noindent
where we redefined some constants
\begin{eqnarray}
\beta\equiv \frac{\kappa_{(5)}^{4}}{36},~~~
\gamma\equiv \frac{\kappa_{(5)}^{4}T_{b}}{36},~~~
\lambda\equiv \frac{1}{3}
\left(\Lambda_{(5)}+\frac{\kappa_{(5)}^{4}T_{b}^{2}}{6}\right).
\label{eq:Chi1b}
\end{eqnarray}

\noindent
Then (\ref{eq:Eeb5}) is rewritten as a pair of the first order equations,
\begin{eqnarray}
\left(\frac{\dot{a}_{b}}{a_{b}}\right)^{2} +\frac{K}{a_{b}^{2}}
&=&
2\gamma\rho_{b} +\beta\rho_{b}^{2} +\chi +\frac{\lambda}{2}
-\frac{\kappa_{(5)}^{2}}{6}P_{T},
\label{eq:Chi3}\\
\dot{\chi}
&=&
-4\frac{\dot{a}_{b}}{a_{b}}\chi +4Q(\gamma+\beta\rho_{b})
+\frac{\kappa_{(5)}^{2}}{6}\dot{P}_{T}.
\label{eq:Chi4}
\end{eqnarray}

\noindent
The equation (\ref{eq:Chi3}) is analogous to the Hubble equation
of standard four dimensional cosmology. However the quadratic term of brane
energy density $\rho_{b}$, the bulk pressure and the brane tension
appear in the right hand side.
The dark energy variable $\chi(t)$ accounts for the non-standard
contributions to the Friedmann equation (\ref{eq:Chi3}).
The evolution of $\rho_{b}$ is determined by (\ref{eq:Ebr1}) and
it is solved as
\begin{eqnarray}
\rho_{b}=
-\left(\frac{a_{b}(0)}{a_{b}(t)}\right)^{3(1+w_{b})}
\int 2Q(t) \left(\frac{a_{b}(t)}{a_{b}(0)}\right)^{3(1+w_{b})} dt.
\label{eq:Chi6}
\end{eqnarray}

\noindent
We treat $w_{b}$ as a constant in the above expressions.
Now we perform the integration of (\ref{eq:Chi4}).
With the ans\"{a}tze (\ref{eq:Chi7}), $\rho_{b}$ is expressed as,
\begin{eqnarray}
\rho_{b}=
\widetilde{\mathcal{C}} a_{b}(t)^{-3(1+w_{b})}
-\frac{2F}{3(1+w_{b})+\mu}a_{b}(t)^{\mu},
\label{eq:Chi8}
\end{eqnarray}

\noindent
where $\widetilde{\mathcal{C}}$ is some integration constant.
With the ans\"{a}tze (\ref{eq:Chi7}), we can perform the integration of
(\ref{eq:Chi4}) and we obtain
\begin{eqnarray}
\chi&=&
\frac{4F\gamma}{4+\mu}a_{b}^{\mu}
+\frac{4F\beta\widetilde{\mathcal{C}}}{\mu-3w_{b}+1}a_{b}^{\mu-3(1+w_{b})}
-\frac{4F^{2}\beta}{(3(1+w_{b})+\mu)(2+\mu)}a_{b}^{2\mu}
\nonumber\\
&&
+\frac{\kappa_{(5)}^{2}D\nu}{6(4+\nu)}a_{b}^{\nu}
+\frac{\mathcal{C}}{a_{b}^{4}},
\label{eq:Chi10}
\end{eqnarray}

\noindent
where $\mathcal{C}$ is some integration constant. Substituting
(\ref{eq:Chi8}) and (\ref{eq:Chi10}) into (\ref{eq:Chi3}), we obtain
the Hubble equation on the brane,
\begin{eqnarray}
\left(\frac{\dot{a}_{b}}{a_{b}}\right)^{2}
&=&
\frac{\lambda}{2}
-\frac{K}{a_{b}^{2}}
+\frac{\mathcal{C}}{a_{b}^{4}}
-\frac{2\kappa_{(5)}^{2}D}{3(4+\nu)}a_{b}^{\nu}
+\frac{2\gamma\widetilde{\mathcal{C}}}{a_{b}^{3(1+w_{b})}}
+\frac{\beta\widetilde{\mathcal{C}}^{2}}{a_{b}^{6(1+w_{b})}}
\nonumber\\
&&
-\frac{4F\gamma(1-3w_{b})}{(4+\mu)(3(1+w_{b})+\mu)}a_{b}^{\mu}
-\frac{4F^{2}\beta(1+3w_{b})}{(3(1+w_{b})+\mu)^{2}(2+\mu)}a_{b}^{2\mu}
\nonumber\\
&&
+\frac{8F\beta\widetilde{\mathcal{C}}(1+3w_{b})}
{(1-3w_{b}+\mu)(3(1+w_{b})+\mu)}a_{b}^{\mu-3(1+3w_{b})}.
\label{eq:Hub1}
\end{eqnarray}

\noindent
This equation is quite different from the Hubble equation
of a standard four dimensional FLRW cosmology:

\vspace*{-0.25\baselineskip}
\begin{list}{$\bullet$}{\itemsep=4pt \parsep=0pt \leftmargin=2em}
\item
The brane energy density appears in a linear and quadratic form, whereas
it appears in a linear form in the standard four dimensional FLRW cosmology.

\item
There is a bulk pressure term $Da_{b}^{\nu}$ and an energy exchange
term $Fa_{b}^{\mu}$ between the four dimensional spacetime and the extra
dimension.

\item
There is a bulk radiation term $\mathcal{C}/a_{b}^{4}$.
\end{list}

\noindent
We can write the Hubble equation (\ref{eq:Hub1}) in the conventional form,
\begin{eqnarray}
\left(\frac{\dot{a}_{b}}{a_{b}}\right)^{2}
&=&
-\frac{K}{a_{b}^{2}}
+\frac{\Lambda_{(4)}}{3}
+\frac{8\pi G_{(4)}}{3}\rho_{eff},
\label{eq:NoF8}\\
\rho_{eff}
&\equiv&
\rho_{b} +\frac{\beta}{2\gamma}\rho_{b}^{2} +\frac{\chi}{2\gamma}
-\frac{\kappa_{(5)}^{2}}{12\gamma}P_{T},
\label{eq:NoF10}
\end{eqnarray}

\noindent
where $\rho_{eff}$ is the effective energy density.
The four dimensional Newton constant $G_{(4)}$ and the four dimensional
cosmological constant $\Lambda_{(4)}$ are defined as
\begin{eqnarray}
G_{(4)}&\equiv&\frac{3\gamma}{4\pi}~=~\frac{4\pi G_{(5)}^{2}T_{b}}{3},
\label{eq:NoF9}\\
\Lambda_{(4)}&\equiv&\frac{3}{2}\lambda~=~
\frac{1}{2}\left(\Lambda_{(5)}+\frac{\kappa_{(5)}^{4}T_{b}^{2}}{6}\right).
\label{eq:NoF9b}
\end{eqnarray}

\noindent
Deceleration parameter $q$ is defined by
\begin{eqnarray}
q&\equiv&
-\left(\frac{a_{b}}{\dot{a}_{b}}\right)^{2}
\left(\frac{\ddot{a}_{b}}{a_{b}}\right),
\label{eq:Acc1}
\end{eqnarray}

\noindent
where the acceleration behavior is described by
\begin{eqnarray}
\frac{\ddot{a}_{b}}{a_{b}}
&=&
\frac{\lambda}{2}
-\frac{\mathcal{C}}{a_{b}^{4}}
+\frac{(2+\nu)\kappa_{(5)}^{2}D}{3(4+\nu)}a_{b}^{\nu}
-\frac{(1+3w_{b})\gamma\widetilde{\mathcal{C}}}{a_{b}^{3(1+w_{b})}}
-\frac{(2+3w_{b})\beta\widetilde{\mathcal{C}}^{2}}{a_{b}^{6(1+w_{b})}}
\nonumber\\
&&
+\frac{2F\gamma(-2-\mu+6w_{b}+3w_{b}\mu)}{(4+\mu)(3(1+w_{b})+\mu)}
a_{b}^{\mu}
-\frac{4F^{2}\beta(1+3w_{b})(1+\mu)}{(3(1+w_{b})+\mu)^{2}(2+\mu)}
a_{b}^{2\mu}
\nonumber\\
&&
-\frac{4F\widetilde{\mathcal{C}}\beta(1+3w_{b})(1+3w_{b}-\mu)}
{(3(1+w_{b})+\mu)(1-3w_{b}+\mu)}
a_{b}^{\mu-3(1+w_{b})}.
\label{eq:Acc2}
\end{eqnarray}

We assume that $K=0$, $\Lambda_{(4)}=0$ and the matter on the brane is all
ordinary non-relativistic matter which is taken to be $w_{b}=0$, and
we obtain
\begin{eqnarray}
H(t)^{2}~=~
\left(\frac{\dot{a}_{b}}{a_{b}}\right)^{2}
&=&
\frac{\mathcal{C}}{a_{b}^{4}}
-\frac{2\kappa_{(5)}^{2}D}{3(4+\nu)}a_{b}^{\nu}
+\frac{2\gamma\widetilde{\mathcal{C}}}{a_{b}^{3}}
+\frac{\beta\widetilde{\mathcal{C}}^{2}}{a_{b}^{6}}
-\frac{4F\gamma}{(4+\mu)(3+\mu)}a_{b}^{\mu}
\nonumber\\
&&
-\frac{4F^{2}\beta}{(3+\mu)^{2}(2+\mu)}a_{b}^{2\mu}
+\frac{8F\beta\widetilde{\mathcal{C}}}
{(1+\mu)(3+\mu)}a_{b}^{\mu-3}.
\label{eq:Hub2}
\end{eqnarray}

\noindent
In the above equation, $H(t)$ gives the expansion rate of the four dimensional
universe. Within the flat universe, in the presence of dark energy,
the expansion rate is given as
\begin{eqnarray}
\frac{H^{2}(t)}{H_{0}^{2}}
=\frac{\Omega_{m}}{a_{b}^{3}}+\frac{1-\Omega_{m}}{a_{b}^{3(1+w)}},
\label{eq:Hub3}
\end{eqnarray}

\noindent
where $H_{0}^{2}\equiv H^{2}(0)$, $\Omega_{m}$ is the dimensionless matter
density and $w$ is the parameter of equation of state of the dark energy.
The dimensionless dark energy density is given by $1-\Omega_{m}$.
The second term in the right hand side of (\ref{eq:Hub3})
describes the contribution of the dark energy to the expansion rate of
our universe.
Following Linder et al.~\cite{Linder}, we modify this equation as
\begin{eqnarray}
\frac{\delta H^{2}}{H_{0}^{2}}&\equiv&
\frac{H^{2}(t)}{H_{0}^{2}}-\frac{\Omega_{m}}{a_{b}^{3}},
\label{eq:NoF12}\\
w_{eff}&\equiv&-1-\frac{1}{3}\frac{d\ln (\delta H^{2})}{d\ln a_{b}},
\label{eq:NoF11}
\end{eqnarray}

\noindent
where $\delta H^{2}/H_{0}^{2}$ accounts for any modification
to the usual Hubble equation in the four dimensional spacetime
and $w_{eff}$ is the effective equation of state parameter.
In our case, $w_{eff}$ is given as
\begin{eqnarray}
w_{eff}(z)
&=&
-1
-\frac{1}{3}
\Biggl(
-4\mathcal{C}(z+1)^{4}
-\frac{2\kappa_{(5)}^{2}\nu D}{3(4+\nu)}(z+1)^{-\nu}
-6\beta\widetilde{\mathcal{C}}^{2}(z+1)^{6}
\nonumber\\
&&
-\frac{4\gamma\mu F}{(4+\mu)(3+\mu)}(z+1)^{-\mu}
-\frac{8\beta\mu F^{2}}{(3+\mu)^{2}(2+\mu)}(z+1)^{-2\mu}
\nonumber\\
&&
+\frac{8\beta(\mu-3)\widetilde{\mathcal{C}}F}{(1+\mu)(3+\mu)}(z+1)^{3-\mu}
\Biggr)
\nonumber\\
&&
\Bigg/
\Biggl(
\mathcal{C}(z+1)^{4}
-\frac{2\kappa_{(5)}^{2}D}{3(4+\nu)}(z+1)^{-\nu}
+\beta\widetilde{\mathcal{C}}^{2}(z+1)^{6}
\nonumber\\
&&
-\frac{4\gamma F}{(4+\mu)(3+\mu)}(z+1)^{-\mu}
-\frac{4\beta F^{2}}{(3+\mu)^{2}(2+\mu)}(z+1)^{-2\mu}
\nonumber\\
&&
+\frac{8\beta\widetilde{\mathcal{C}}F}{(1+\mu)(3+\mu)}(z+1)^{3-\mu}
\Biggr),
\label{eq:withF2b}
\end{eqnarray}

\noindent
where we use the redshift parameter $z$ defined as
$z(t)+1\equiv a(t_{0})/a(t)$. Deceleration parameter is given by
\begin{eqnarray}
q(z)
&=&
-\Biggl(
-\mathcal{C}(z+1)^{4}
+\frac{(2+\nu)\kappa_{(5)}^{2}D}{3(4+\nu)}(z+1)^{-\nu}
-\gamma\widetilde{\mathcal{C}}(z+1)^{3}
-2\beta\widetilde{\mathcal{C}}^{2}(z+1)^{6}
\nonumber\\
&&
-\frac{2F\gamma(2+\mu)}{(4+\mu)(3+\mu)}(z+1)^{-\mu}
-\frac{4F^{2}\beta(1+\mu)}{(3+\mu)^{2}(2+\mu)}(z+1)^{-2\mu}
\nonumber\\
&&
-\frac{4F\widetilde{\mathcal{C}}\beta(1-\mu)}{(3+\mu)(1+\mu)}(z+1)^{3-\mu}
\Biggr)
\nonumber\\
&&
\Bigg/
\Biggl(
\mathcal{C}(z+1)^{4}
-\frac{2\kappa_{(5)}^{2}D}{3(4+\nu)}(z+1)^{-\nu}
+2\gamma\widetilde{\mathcal{C}}(z+1)^{3}
+\beta\widetilde{\mathcal{C}}^{2}(z+1)^{6}
\nonumber\\
&&
-\frac{4F\gamma}{(4+\mu)(3+\mu)}(z+1)^{-\mu}
-\frac{4F^{2}\beta}{(3+\mu)^{2}(2+\mu)}(z+1)^{-2\mu}
\nonumber\\
&&
+\frac{8F\beta\widetilde{\mathcal{C}}}{(1+\mu)(3+\mu)}(z+1)^{3-\mu}
\Biggr).
\label{eq:withF3b}
\end{eqnarray}

\section{Evolution of $w_{eff}$ and the crossing $w_{eff}=-1$}

We examine the evolution of the effective equation of state parameter
$w_{eff}$ and investigate under which condition it exhibits $w_{eff}<-1$.
The cosmological data indicates that
$w_{eff}\sim -1.21$ at $z=0$ and it changes
from $w_{eff}>-1$ to $w_{eff}<-1$ at $z\sim 0.2$~\cite{Alam}.
We look for the set of parameters which exhibit such behavior
with the assumptions that $K=0$ and $\lambda=0$.
Each parameters must satisfy the following constraint which is obtained
from (\ref{eq:Hub2}),
\begin{eqnarray}
&&
\frac{\mathcal{C}}{H^{2}_{0}}
-\frac{2\kappa_{(5)}^{2}D}{3(4+\nu)H^{2}_{0}}
-\frac{4F\gamma}{(4+\mu)(3+\mu)H^{2}_{0}}
\nonumber\\
&&
-\frac{4F^{2}\beta}{(3+\mu)^{2}(2+\mu)H^{2}_{0}}
+\frac{8F\beta\widetilde{\mathcal{C}}}{(1+\mu)(3+\mu)H^{2}_{0}}
+\frac{2\gamma\widetilde{\mathcal{C}}}{H^{2}_{0}}
+\frac{\beta\widetilde{\mathcal{C}}^{2}}{H^{2}_{0}}
~=~1.
\label{eq:withF1b}
\end{eqnarray}

\noindent
In the above expression, $2\gamma\widetilde{\mathcal{C}}/H^{2}_{0}$
corresponds to the dimensionless matter density and we assume that
$2\gamma\widetilde{\mathcal{C}}/H^{2}_{0}<1$.
We consider two cases:

\vspace*{-0.25\baselineskip}
{\newcommand{\Ai}{\refstepcounter{Ai}(\roman{Ai})}
\setcounter{Ai}{0}
\begin{list}{\Ai}{\itemsep=4pt \parsep=0pt \leftmargin=2em}
\item
there is energy exchange between the four dimensional universe and
the fifth dimension,

\item
there is no energy exchange between the four dimensional universe and
the fifth dimension.
\end{list}}

\subsection{No quadratic brane energy density and no bulk radiation}

We consider that there is energy exchange between the four dimensional
universe and the fifth dimension.
We assume that the brane energy density is much smaller
than the brane tension: $\rho_{b}\ll T_{b}$.
In this case we can neglect the quadratic term in $\rho_{b}$.
We also assume that the bulk radiation is negligible: $\mathcal{C}=0$.
We still assume that the brane matter is all ordinary matter ($w_{b}=0$)
and we use the ansatz (\ref{eq:Chi7}).
With these assumptions, we obtain the Hubble equation (\ref{eq:Hub1})
and the acceleration behavior (\ref{eq:Acc2}) as follows,
\begin{eqnarray}
\left(\frac{\dot{a}_{b}}{a_{b}}\right)^{2}
&=&
-Aa_{b}^{\nu}-Ba_{b}^{\mu}+\frac{C}{a_{b}^{3}},
\label{eq:No-rho2C-1a}\\
\frac{\ddot{a}_{b}}{a_{b}}
&=&
\frac{(2+\nu)A}{2}a_{b}^{\nu}-\frac{(2+\mu)B}{2}a_{b}^{\mu}
-\frac{C}{2a_{b}^{3}},
\label{eq:No-rho2C-1b}
\end{eqnarray}

\noindent
where we used the notations
\begin{eqnarray}
A\equiv
\frac{2\kappa_{(5)}^{2}w_{B}C_{B}}{3(4+\nu)},~
B\equiv
\frac{4F\gamma}{(4+\mu)(3+\mu)},~
C\equiv
2\gamma\widetilde{\mathcal{C}}.
\label{eq:No-rho2C-2c}
\end{eqnarray}

\noindent
The parameter $A$ corresponds to the contribution to Hubble parameter
from the bulk matter, $B$ corresponds to the contribution
from the the energy exchange between the extra dimension and
$C$ corresponds to the contribution from the ordinary matter on the brane.
The Hubble equation (\ref{eq:No-rho2C-1a}) is expressed as,
\begin{eqnarray}
-A-B+C&=&H_{0}^{2}.
\label{eq:No-rho2C-2b}
\end{eqnarray}

\noindent
The first two terms correspond to effective dark energy density.

The deceleration parameter $q$, the effective equation of state parameter
$w_{eff}$ and its present value are given by
\begin{eqnarray}
q
&=&
\frac{\displaystyle
\frac{(2+\nu)A}{2(z+1)^{\nu}}
-\frac{(2+\mu)B}{2(z+1)^{\mu}}
-\frac{C}{2}(z+1)^{3}
}
{\displaystyle
\frac{A}{(z+1)^{\nu}}
+\frac{B}{(z+1)^{\mu}}
-C(z+1)^{3}
},
\label{eq:No-rho2C-5}\\
w_{eff}
&=&
-1-\frac{1}{3}\left(
\frac{\displaystyle
\frac{\nu A}{(z+1)^{\nu}}
+\frac{\mu B}{(z+1)^{\mu}}
}
{\displaystyle
\frac{A}{(z+1)^{\nu}}
+\frac{B}{(z+1)^{\mu}}
}
\right),
\label{eq:No-rho2C-3}\\
w_{eff}(0)
&=&
-1-\frac{1}{3}\left(
\frac{\nu A+\mu B}{A+B}
\right).
\label{eq:No-rho2C-4}
\end{eqnarray}

\noindent
We look for the set of parameters which satisfy
$w_{eff}<-1$ for $0<z<0.2$ and $w_{eff}>-1$ for $z<0.2$.
The denominators of (\ref{eq:No-rho2C-3}) and (\ref{eq:No-rho2C-4})
correspond to the opposite sign of effective dark energy density and
it must be negative. We can achieve $w_{eff}<-1$ when their numerators
are negative. Then we obtain two conditions,
\begin{eqnarray}
A\nu (z+1)^{-\nu} +B\mu (z+1)^{-\mu} &<& 0,
\label{eq:No-rho2C-4b}\\
A(z+1)^{-\nu} +B(z+1)^{-\mu} &<& 0.
\label{eq:No-rho2C-4c}
\end{eqnarray}

\noindent
for $z<0.2$. We can achieve $w_{eff}(0)=-1.21$ when
\begin{eqnarray}
A(\nu-0.63)+B(\mu-0.63)=0.
\label{eq:No-rho2C-4d}
\end{eqnarray}

\noindent
There is another constraint which comes from $w_{eff}(0.2)=-1$,
\begin{eqnarray}
\frac{\nu}{\mu}
=-\frac{B}{A}(1.2)^{\nu-\mu}.
\label{eq:No-rho2C-6}
\end{eqnarray}

\noindent
This equation determines the relative sign of $\mu$ and $\nu$.
The allowed choices of parameters are found in~\cite{Bogdanos-b} and they are:

\vspace*{-0.25\baselineskip}
{\newcommand{\Ai}{\refstepcounter{Ai}(\roman{Ai})}
\setcounter{Ai}{0}
\begin{list}{\Ai}{\itemsep=4pt \parsep=0pt \leftmargin=2em}
\item
$A,B<0$ when $\mu >0, \nu <0$,

\item
$A,B<0$ when $\mu <0, \nu >0$.
\end{list}}

\begin{figure}[ht]
\begin{tabular}{cc}
\begin{minipage}{75mm}
\begin{center}
\epsfig{file=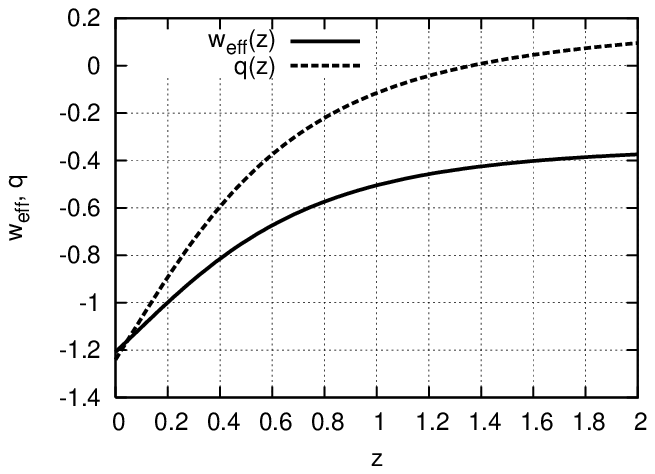,width=72mm}
\end{center}
\vspace{-\baselineskip}
\caption{\small Graph of $w_{eff}(z)$ and $q(z)$ with parameters
$A=-0.30$, $B=-0.66$, $C=0.04$, $\mu=1.82$, $\nu=-2$.}
\label{fig:No-rho2C-1}
\end{minipage}&
\begin{minipage}{75mm}
\begin{center}
\epsfig{file=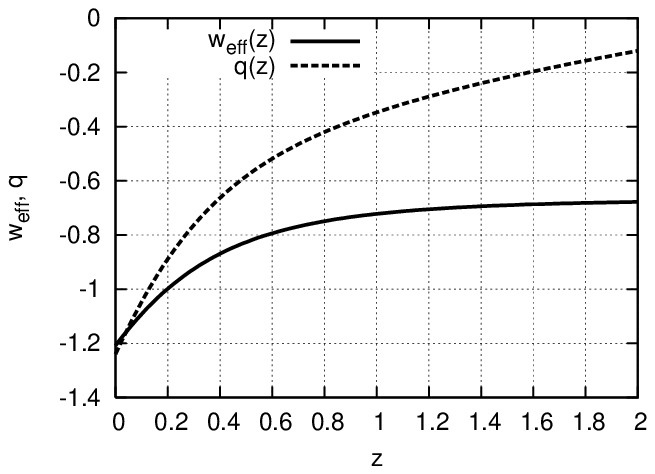,width=72mm}
\end{center}
\vspace{-\baselineskip}
\caption{\small Graph of $w_{eff}(z)$ and $q(z)$ with parameters
$A=-0.57$, $B=-0.39$, $C=0.04$, $\mu=2.98$, $\nu=-1$.}
\label{fig:No-rho2C-2}
\end{minipage}
\end{tabular}
\end{figure}

We look for the set of parameters which satisfy (\ref{eq:No-rho2C-4d})
and (\ref{eq:No-rho2C-6}) under the constraint of (\ref{eq:No-rho2C-2b}).
When we assume $\nu=-2$ and $C=0.04$, we obtain $A=-0.30$, $B=-0.66$ and
$\mu=1.82$. Figure \ref{fig:No-rho2C-1} show the behaviors of $w_{eff}(z)$
and $q(z)$ in this case.
When we assume $\nu=-1$ and $C=0.04$, we obtain $A=-0.57$, $B=-0.39$ and
$\mu=2.98$. Figure \ref{fig:No-rho2C-2} show the behaviors of $w_{eff}(z)$
and $q(z)$ in this case.
In each cases, $w_{eff}(z)$ and $q(z)$ increase with $z$ and $q(z)$
become positive $z\simeq 1.4$ in figure \ref{fig:No-rho2C-1} and
$z\simeq 2.8$ in figure \ref{fig:No-rho2C-2}.
The behavior of $w_{eff}(z)$ and $q(z)$ with the parameters
$\left( A,B,C,\mu,\nu \right)=(-1,-2,-2,2,-2)$ is found in~\cite{Bogdanos-b}.

We consider the behavior of energy density components.
We write the constraint (\ref{eq:No-rho2C-1a}) as
\begin{eqnarray}
\Omega_{A}(z)+\Omega_{B}(z)+\Omega_{C}(z)=1,
\label{eq:No-rho2C-7}
\end{eqnarray}

\noindent
where
\begin{eqnarray}
\Omega_{A}(z)\equiv
-\frac{A}{(z+1)^{\nu}H^{2}(z)},~~
\Omega_{B}(z)\equiv
-\frac{B}{(z+1)^{\mu}H^{2}(z)},~~
\Omega_{C}(z)\equiv
\frac{C(z+1)^{3}}{H^{2}(z)}.
\label{eq:No-rho2C-8}
\end{eqnarray}

\begin{figure}[ht]
\begin{tabular}{cc}
\begin{minipage}{75mm}
\begin{center}
\epsfig{file=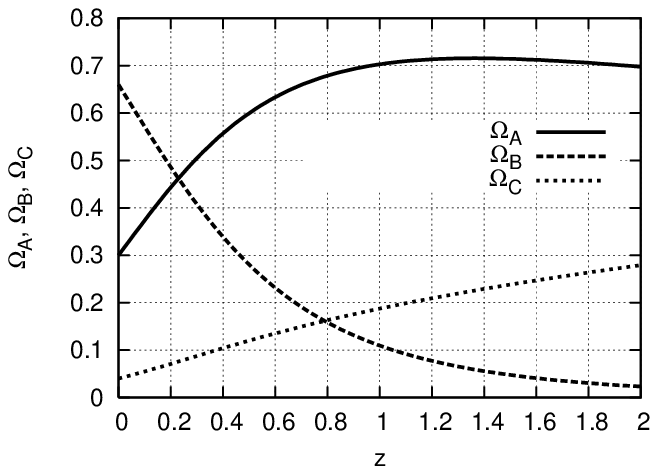,width=72mm}
\end{center}
\vspace{-\baselineskip}
\caption{\small Graph of $\Omega_{A},\,\Omega_{B},\,\Omega_{C}$
with parameters $A=-0.30$, $B=-0.66$, $C=0.04$, $\mu=1.82$, $\nu=-2$.}
\label{fig:No-rho2C-3}
\end{minipage}&
\begin{minipage}{75mm}
\begin{center}
\epsfig{file=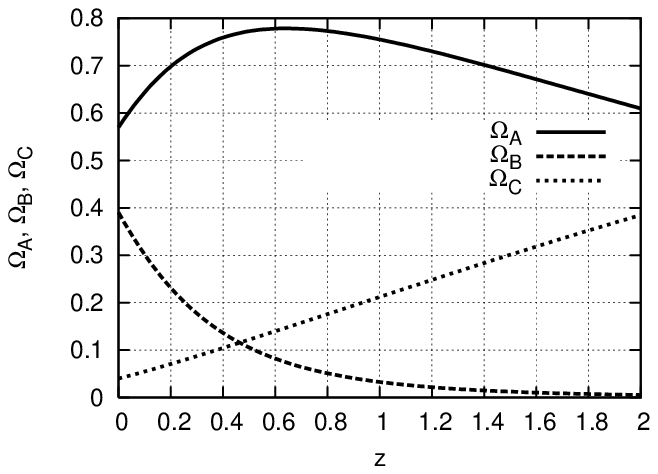,width=72mm}
\end{center}
\vspace{-\baselineskip}
\caption{\small Graph of $\Omega_{A},\,\Omega_{B},\,\Omega_{C}$
with parameters $A=-0.57$, $B=-0.39$, $C=0.04$, $\mu=2.98$, $\nu=-1$.}
\label{fig:No-rho2C-4}
\end{minipage}
\end{tabular}
\end{figure}

\noindent
$\Omega_{A}$ corresponds to the contribution from the bulk matter,
$\Omega_{B}$ corresponds to the contribution from the energy exchange
between the extra dimension and $\Omega_{C}$ corresponds to the contribution
from the brane matter.
The behavior of each energy density components are shown in figures
\ref{fig:No-rho2C-3} and \ref{fig:No-rho2C-4}.
In figure \ref{fig:No-rho2C-3}, $\Omega_{B}$ is dominant in small $z$ and
$\Omega_{A}$ is dominant in large $z$.
In figure \ref{fig:No-rho2C-4}, $\Omega_{A}$ is dominant even in small $z$.
The linear contributions from the brane matter $\Omega_{C}$ are smaller
than the other part, but they increase with $z$.
$\Omega_{C}$ become subdominant at $z\sim 0.8$ in figure
\ref{fig:No-rho2C-3} and $z\sim 0.4$ in figure \ref{fig:No-rho2C-4}.

\subsection{No energy flow to/from the extra dimension}

We assume that there is no energy flow to/from the extra dimension:
$T^{0}{}_{5}=T^{5}{}_{0}=0$. It corresponds to $F=0$.
We put $F=0$ and $P_{T}=Da_{b}^{\nu}$ in (\ref{eq:Chi8})
and (\ref{eq:Chi10}), we obtain
\begin{eqnarray}
\rho_{b}&=&
\frac{\widetilde{\mathcal{C}}}{a_{b}^{3(1+w_{b})}},
\label{eq:NoF2}\\
\chi&=&
\frac{\mathcal{C}}{a_{b}^{4}}
+\frac{\kappa_{(5)}^{2}D\nu}{6(4+\nu)}a_{b}^{\nu},
\label{eq:NoF4}
\end{eqnarray}

\noindent
where $\mathcal{C}$ and $\widetilde{\mathcal{C}}$ are some constants.
The equation for the bulk matter (\ref{eq:Ebu1}) is expressed as,
\begin{eqnarray}
\dot{\rho}_{B}&=&
-3(P_{B}+\rho_{B})\frac{\dot{a}_{b}}{a_{b}},
\label{eq:NoF16}
\end{eqnarray}

\noindent
where we assumed $P_{B}=w_{B}\rho_{B}$ for the bulk matter.
It is solved as
\begin{eqnarray}
\rho_{B}&=&\frac{C_{B}}{a_{b}^{3(1+w_{B})}},
\label{eq:NoF17}
\end{eqnarray}

\noindent
where $C_{B}$ is some integration constant. We obtain the relation
\begin{eqnarray}
P_{T}=P_{B}=\frac{w_{B}C_{B}}{a_{b}^{3(1+w_{B})}},
\label{eq:NoF18}
\end{eqnarray}

\noindent
when we assume that the pressure of bulk matter is isotropic.
We read $\nu=-3(1+w_{B})$ and $D=w_{B}C_{B}$ from (\ref{eq:Chi7}).
Assuming $w_{b}=0$,
the Hubble equation and the acceleration behavior is expressed as follows,
\begin{eqnarray}
\left(\frac{\dot{a}_{b}}{a_{b}}\right)^{2}
&=&
\frac{A}{a_{b}^{4}}
-\frac{B}{a_{b}^{3(1+w_{B})}}
+\frac{C}{a_{b}^{3}}
+\frac{D}{a_{b}^{6}},
\label{eq:NoF5}\\
\frac{\ddot{a}_{b}}{a_{b}}
&=&
-\frac{A}{a_{b}^{4}}
-\frac{B(1+3w_{B})}{2a_{b}^{3(1+w_{B})}}
-\frac{C}{2a_{b}^{3}}
-\frac{2D}{a_{b}^{6}}.
\label{eq:NoF6}
\end{eqnarray}

\noindent
where we used the notation
\begin{eqnarray}
A\equiv \mathcal{C},~
B\equiv \frac{2\kappa_{(5)}^{2}w_{B}C_{B}}{3(1-3w_{B})},~
C\equiv 2\gamma\widetilde{\mathcal{C}},~
D\equiv \beta\widetilde{\mathcal{C}}^{2}.
\label{eq:NoF6b}
\end{eqnarray}

\noindent
Each parameters must satisfy the constraint from (\ref{eq:NoF5}),
\begin{eqnarray}
A-B+C+D&=&H_{0}^{2}
\label{eq:NoF6c}
\end{eqnarray}

\noindent
where $C$ corresponds to the energy density of ordinary matter.
The deceleration parameter $q$, the effective equation of state parameter
$w_{eff}$ and its present value are given by
\begin{eqnarray}
q
&=&
\Biggl(
A(1+z)^{4}
+\frac{(1+3w_{B})B}{2}(1+z)^{3(1+w_{B})}
+\frac{C}{2}(1+z)^{3}
+2D(1+z)^{6}
\Biggr)
\nonumber\\
&&
\Bigg/
\Biggl(
A(1+z)^{4}
-B(1+z)^{3(1+w_{B})}
+C(1+z)^{3}
+D(1+z)^{6}
\Biggr),
\label{eq:NoF20}\\
w_{eff}
&=&
-1+\frac{1}{3}\Biggl\{
4A(z+1)^{4}
-3B(1+w_{B})(z+1)^{3(1+w_{B})}
+6D(z+1)^{6}
\Biggr\}
\nonumber\\
&&
\Bigg/
\Biggl\{
A(z+1)^{4}
-B(z+1)^{3(1+w_{B})}
+D(z+1)^{6}
\Biggr\},
\label{eq:NoF19}\\
w_{eff}(0)
&=&
-1+\frac{1}{3}
\left(
\frac{4A-3B(1+w_{B})+6D}{A-B+D}
\right).
\label{eq:NoF25}
\end{eqnarray}

We look for the set of parameters which satisfy
$w_{eff}<-1$ for $0<z<0.2$ and $w_{eff}>-1$ for $z<0.2$.
The denominators of (\ref{eq:NoF19}) and (\ref{eq:NoF25}) correspond to
the effective dark energy density and it must be positive.
We can achieve $w_{eff}<-1$ when their numerators are negative.
Then we obtain two conditions for $z<0.2$,
\begin{eqnarray}
4A(z+1)^{4} -3B(1+w_{B})(z+1)^{3(1+w_{B})} +6D(z+1)^{6} &<& 0,
\label{eq:NoF25b}\\
A(z+1)^{4} -B(z+1)^{3(1+w_{B})} +D(z+1)^{6} &>& 0.
\label{eq:NoF25c}
\end{eqnarray}

\noindent
They gives the upper and the lower bound to achieve $w_{eff}(0)=-1.21$.
We can achieve $w_{eff}(0)=-1.21$ when
\begin{eqnarray}
4.63A-3B(1.21+w_{B})+6.63D=0.
\label{eq:NoF25d}
\end{eqnarray}

\noindent
There is another constraint which comes from $w_{eff}(0.2)=-1$,
\begin{eqnarray}
4A(1.2)^{4}-3B(1+w_{B})(1.2)^{3(1+w_{B})}+6D(1.2)^{6}=0.
\label{eq:NoF25e}
\end{eqnarray}

\begin{figure}[ht]
\begin{tabular}{cc}
\begin{minipage}{75mm}
\begin{center}
\epsfig{file=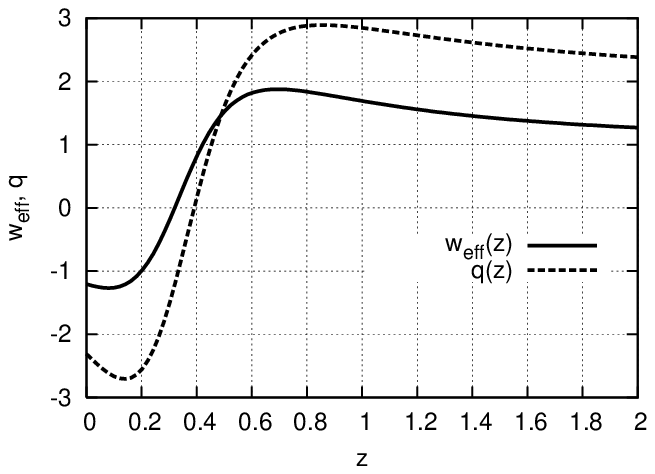,width=72mm}
\end{center}
\vspace{-\baselineskip}
\caption{\small Graph of $w_{eff}(z)$ and $q(z)$ with parameters
$w_{B}=-1/3$, $A=-3.19$, $B=-3.19$, $C=0.04$, $D=0.96$.}
\label{fig:NoF-1}
\end{minipage}&
\begin{minipage}{75mm}
\begin{center}
\epsfig{file=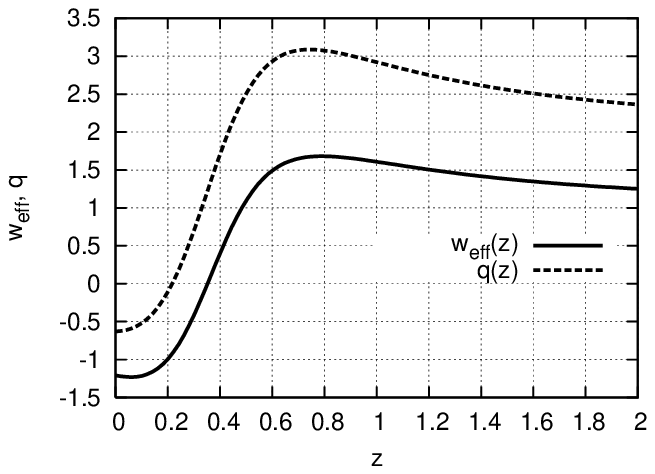,width=72mm}
\end{center}
\vspace{-\baselineskip}
\caption{\small Graph of $w_{eff}(z)$ and $q(z)$ with parameters
$w_{B}=-2/3$, $A=-1.39$, $B=-1.83$, $C=0.04$, $D=0.52$.}
\label{fig:NoF-2}
\end{minipage}
\end{tabular}
\begin{center}
\begin{minipage}{75mm}
\begin{center}
\epsfig{file=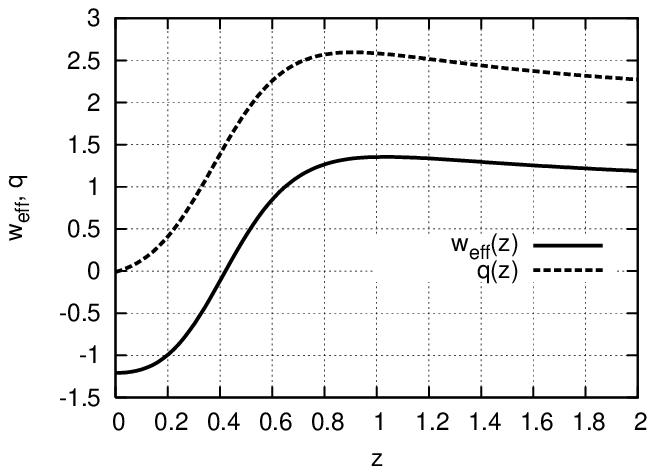,width=72mm}
\end{center}
\vspace{-\baselineskip}
\caption{\small Graph of $w_{eff}(z)$ and $q(z)$ with parameters
$w_{B}=-1$, $A=-0.49$, $B=-1.22$, $C=0.04$, $D=0.23$.}
\label{fig:NoF-3}
\end{minipage}
\end{center}
\end{figure}

\noindent
We look for the set of parameters which satisfy (\ref{eq:NoF25d})
and (\ref{eq:NoF25e}) under the condition of (\ref{eq:NoF6c}).
When we assume $w_{B}=-1/3$ and $C=0.04$, we obtain $A=-3.19$, $B=-3.19$ and
$D=0.96$. Figure \ref{fig:NoF-1} show the behaviors of $w_{eff}(z)$
and $q(z)$ in this case.
When we assume $w_{B}=-2/3$ and $C=0.04$, we obtain $A=-1.39$, $B=-1.83$ and
$D=0.52$. Figure \ref{fig:NoF-2} show the behaviors of $w_{eff}(z)$
and $q(z)$ in this case.
When we assume $w_{B}=-1$ and $C=0.04$, we obtain $A=-0.49$, $B=-1.22$ and
$D=0.23$. Figure \ref{fig:NoF-3} show the behaviors of $w_{eff}(z)$
and $q(z)$ in this case.
In each cases, $q(z)$ become positive $z\simeq 0.4$ in
figure \ref{fig:NoF-1}, $z\simeq 0.2$ in figure \ref{fig:NoF-2} and
$z\simeq 0.02$ in figure \ref{fig:NoF-3}.

We consider the behavior of energy density components.
We write the constraint (\ref{eq:NoF5}) as
\begin{eqnarray}
\Omega_{A}(z)+\Omega_{B}(z)+\Omega_{C}(z)+\Omega_{D}(z)=1,
\label{eq:NoF39b}
\end{eqnarray}

\noindent
where
\begin{eqnarray}
&&\Omega_{A}(z)\equiv
\frac{\mathcal{C}(z+1)^{4}}{H^{2}(z)},~~
\Omega_{B}(z)\equiv
-\frac{2\kappa_{(5)}^{2}w_{B}C_{B}(z+1)^{3(1+w_{B})}}{3(1-3w_{B})H^{2}(z)},
\nonumber\\
&&\Omega_{C}(z)\equiv
\frac{2\gamma\widetilde{\mathcal{C}}(1+z)^{3(1+w_{b})}}{H^{2}(z)},~~
\Omega_{D}(z)\equiv
\frac{\beta\widetilde{\mathcal{C}}^{2}(1+z)^{6(1+w_{b})}}{H^{2}(z)}.
\label{eq:NoF39a}
\end{eqnarray}

\begin{figure}[ht]
\begin{tabular}{cc}
\begin{minipage}{75mm}
\begin{center}
\epsfig{file=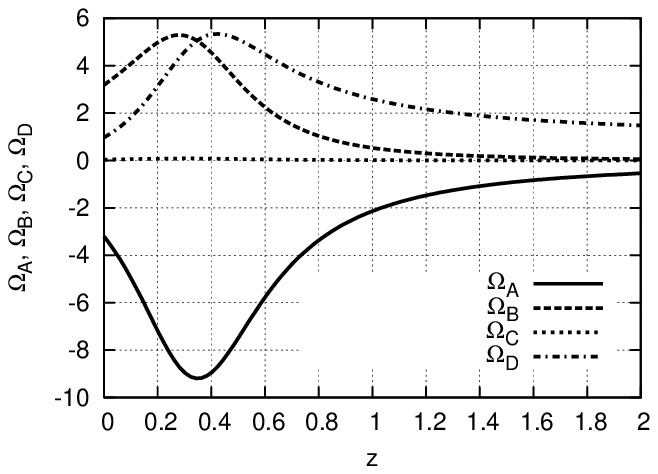,width=72mm}
\end{center}
\vspace{-\baselineskip}
\caption{\small
Graph of $\Omega_{A},\,\Omega_{B},\,\Omega_{C},\,\Omega_{D}$
with parameters $w_{B}=-1/3$, $A=-3.19$, $B=-3.19$, $C=0.04$, $D=0.96$.}
\label{fig:NoF-4}
\end{minipage}&
\begin{minipage}{75mm}
\begin{center}
\epsfig{file=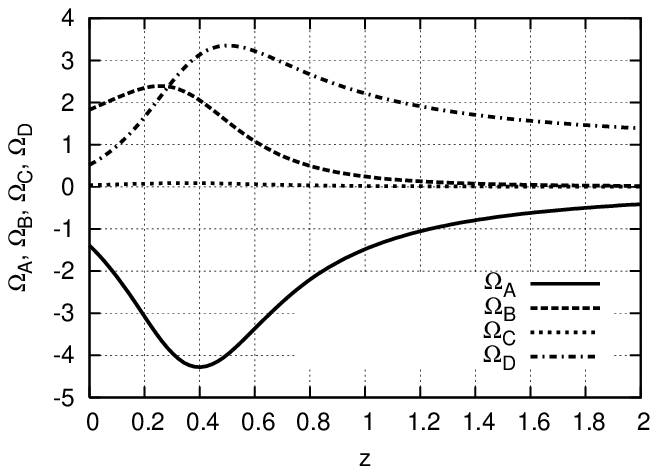,width=72mm}
\end{center}
\vspace{-\baselineskip}
\caption{\small
Graph of $\Omega_{A},\,\Omega_{B},\,\Omega_{C},\,\Omega_{D}$
with parameters $w_{B}=-2/3$, $A=-1.39$, $B=-1.83$, $C=0.04$, $D=0.52$.}
\label{fig:NoF-5}
\end{minipage}
\end{tabular}
\begin{center}
\begin{minipage}{75mm}
\begin{center}
\epsfig{file=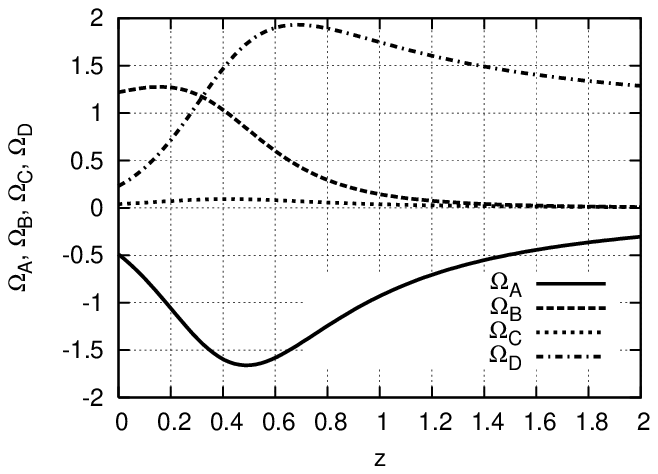,width=72mm}
\end{center}
\vspace{-\baselineskip}
\caption{\small
Graph of $\Omega_{A},\,\Omega_{B},\,\Omega_{C},\,\Omega_{D}$
with parameters $w_{B}=-1$, $A=-0.49$, $B=-1.22$, $C=0.04$, $D=0.23$.}
\label{fig:NoF-6}
\end{minipage}
\end{center}
\end{figure}

\noindent
$\Omega_{A}$ corresponds to the contribution from the bulk radiation,
$\Omega_{B}$ corresponds to the contribution from the bulk matter,
$\Omega_{C}$ corresponds to the linear contribution from the brane matter
and $\Omega_{D}$ corresponds to the quadratic contribution from the brane
matter. The behavior of each energy density components are shown in figures
\ref{fig:NoF-4}--\ref{fig:NoF-6}.

\noindent
There are negative contributions from the bulk radiation to the total
energy density in each cases. The contributions from the bulk matter are
dominant in late time and the quadratic contributions from the brane matter
are dominant in early time. The linear contributions from the brane matter
are much smaller than the other contributions through the time.

\section{Summary}

The cosmological observation indicates that
the effective equation of state parameter $w_{eff}$ varies with $z$:
$w_{eff}\sim -1.21$ at $z=0$ and it crosses $w_{eff}=-1$ at $z\sim 0.2$.
We investigated that under which condition this behavior occurs
based on the five-dimensional braneworld scenario.
The Hubble equation on the 3-brane is quite different from that of
a standard four dimensional FLRW cosmology:
(i) The brane energy density appears in a linear and quadratic form.
(ii) A bulk pressure term, a bulk radiation term, and an energy exchange term
between the four dimensional spacetime and the extra dimension appear
in the Hubble equation.
They contribute to the effective equation of state parameter.

We considered two cases:
(i) There is energy exchange between the four dimensional universe and the
fifth dimension.
(ii) There is no energy exchange between the four dimensional universe and
the fifth dimension.
In both cases,we obtained that the crossing of $w_{eff}=-1$ line
and the universe changes from deceleration to acceleration
at lower redshift. However, the curves of $w_{eff}$ are different
between two cases.
Although it remains negative value in the case (i), it reachs $w_{eff}>1$
in the case (ii).
The curves of $q$ and the energy density components
are also different between two cases.
Especially, there are negative contributions from the bulk radiation to
the total energy density in the case (ii).
In the case (i), the linear contributions from the brane matter is smaller
than the other part, but they increase with $z$.
In the case (ii), they remain much smaller than the other contributions,
however the sum of the linear and quadratic contribution is dominant.


\end{document}